\documentclass[manuscript]{aastex}

\usepackage{cases}

\shorttitle{A New $E_p-L_p$ Relation of Short GRBs}
\shortauthors{Zhang \& Chen}

\begin{document}


\title{Correlation between Peak Energy and Peak Luminosity in Short Gamma-Ray Bursts}


\author{Z. B. Zhang\altaffilmark{1,2}, D. Y. Chen\altaffilmark{1,2} and Y. F. Huang\altaffilmark{3}}
%




\altaffiltext{1}{Department of Physics, College of Sciences, Guizhou
University, Guiyang 550025, China; sci.zbzhang@gzu.edu.cn}
\altaffiltext{2}{The Joint Research Center for Astronomy between National Astronomical
Observatories, CAS and Guizhou University, Guizhou, China}
\altaffiltext{3}{Department of Astronomy, Nanjing University, Nanjing 210093, China; hyf@nju.edu.cn}


\begin{abstract}
A correlation between the peak luminosity and the peak energy has been found by Yonetoku et al. as $L_{p}\propto E_{p,i}^{2.0}$ for 11 pre-Swift long gamma-ray bursts. In this study, for a greatly expanded sample of 148 long gamma-ray bursts in the Swift era, we find that the correlation still exists, but most likely with a slightly different power-law index, i.e., $L_{p}\propto E_{p,i} ^{1.7}$. In addition, we have collected 17 short gamma-ray bursts with necessary data. It is found that the correlation of $L_{p}\propto E_{p,i} ^{1.7}$ also exists for this sample of short events. It is argued that the radiation mechanism of both long and short gamma-ray bursts should be similar, i.e., of quasi-thermal origin caused by the photosphere and the dissipation occurring very near the central engine. Some key parameters of the process are constrained. Our results suggest that the radiation process of both long and short bursts may be dominated by thermal emission, rather than the single synchrotron radiation. This might put strong physical constraints on the theoretical models.
\end{abstract}





\section{Introduction}
Cosmic Gamma-Ray Bursts (GRBs) are the most violent and farthest stellar explosions observed so far. One of their advantages is that GRBs can be used to explore the very early cosmos by virtue of their high cosmological redshifts ($z$). The highest redshift measured in GRBs has now been up to 9.4 (Cucchiara et al. 2011), which makes them a promising tool for cosmological studies. This profits from a number of empirical spectral energy relations discovered during the past decade, i.e., between the spectral lag ($\tau$), the variability ($V$), the relative spectral lag ($\tau_{rel}$), the number of peaks ($N_{peak}$) of GRB light curves, the minimum rise time ($\tau_{RT}$) of GRB pulses, the peak or isotropic luminosity ($L_{p}$ or $L_{iso}$), the rest frame peak energy ($E_{p,i}$), the isotropic energy ($E_{iso}$), and the jet-corrected energy ($E_{\gamma}$). The familiar relations include $\tau-L_{p}$ (Norris, Marani \& Bonnell 2000), $V-L_{p}$ (Fenimore \& Ramirez-Ruiz 2000; Reichart et al. 2001), $E_{p,i}-E_{iso}$ (Amati et al. 2002), $E_{p,i}-E_{\gamma}$ (Ghirlanda, Ghisellini \& Lazzati 2004), $E_{p,i}-L_{p}$ (Yonetoku et al. 2004; see also Wei \& Gao 2003; Schaefer 2003), $N_{peak}-L_{p}$ ( Schaefer 2003), $\tau_{rel}-L_p$ (Zhang et al. 2006, 2008), $\tau_{RT}-L_p$ (Schaefer 2007), etc. It is interesting to note that for a sub-sample of long GRBs with known redshifts and a plateau phase in the afterglow, a tight three-parameter correlation may exist between the end time of the plateau phase (in the GRB rest frame), the corresponding X-ray luminosity and the isotropic $\gamma$-ray energy release (Xu \& Huang 2012).

Among these relations, many efforts have been put on the physical origin and implications of the intriguing $E_{p,i}-L_{p}$ relation (e.g. Kumar \& Piran 2000; Zhang \& M\'{e}sz\'{a}ros 2002; Yamazaki et al. 2004; Eichler \& Levinson 2004; Liang et al. 2004; Rees \& M\'{e}sz\'{a}ros 2005; Toma et al. 2005; Ryde F. et al., 2006; Thompson, M\'{e}sz\'{a}ros \& Rees 2007; Lazzati et al. 2011; Nava et al. 2008, 2010; Ghirlanda et al. 2011a, 2011b). For example, Lloyd-Ronning \& Ramirez-Ruiz (2002) found that bursts with highly variable curves generally have greater spectral peak energies. Since the light curve variability may be correlated with luminosity, it is then reasonable that $E_{p,i}$ and $L_p$ should be correlated. However, the physical mechanisms leading to these correlations are still quite unclear. Some authors even argue that at least several of the above relations, even including the $E_{p,i}-L_{p}$ relation, could be resulted from selection effects (e.g. Band \& Preece 2005; Nakar \& Piran 2005; Butler et al. 2007, 2009; Shahmoradi \& Nemiroff 2010). Therefore, common views are still NOT obtained for the real physical origin of the $E_{p,i}-L_{p}$ relation (Ghirlanda et al. 2011a, 2011b).

 According to their durations ($T_{90}$), GRBs are generally divided into two groups of long and short GRBs (Kouveliotou 1993) at the dividing point of 2 seconds, which is also confirmed by Zhang \& Choi (2008) for Swift bursts with known redshifts. Note that many of the empirical relations requiring for redshift measurements have been discovered and applied only for long GRBs. Application of short GRBs with lower redshifts (Gehrels, Ramirez-Ruiz, Fox 2009; this work) on cosmological studies is an essential complement to long GRBs. Especially, it is very interesting to check whether a similar $E_{p,i}-L_{p}$ relation also exists in short GRBs. Since 2005, statistical investigations of the $E_{p,i}-L_{p}$ relation of short GRBs become possible and more and more prospective because of the accumulation of short-hard GRBs with measured redshifts. Recently, Swift and Fermi long GRBs are found to follow the $E_{p,i}-L_p$ relation within individual bursts and/or a sample of GRBs (e.g. Firmani et al., 2009; Ghirlanda, Nava \& Ghisellini 2010). By calculating the time-resolved spectra of five long Fermi GRBs (080916C, 090424, 090618, 090902, 090926) and one short event (GRB 090510), Ghirlanda, Ghisellini \& Nava (2011b) found that the $E_{p,i}-L_{p}$ relation also existed within individual bursts and that the short burst GRB 090510 was marginally located at the end of long GRB population, which was argued to be the evidence supporting the intrinsic origin of this relation (see also Ghirlanda et al. 2012). In addition, some recent works qualitatively draw a conclusion that short and long bursts should follow the same $E_{p,i}-L_p$ relation (Ghirlanda et al. 2009; Zhang et al. 2009; Zhang et al. 2012). This conclusion is different from the cases of $E_{p}-E_{iso}$ and $E_{p}-E_{\gamma}$  relations, which are derived from long bursts, but the majority of short bursts do not follow (Amati 2006, 2009; Ghirlanda et al. 2009).

Motivated by above situations, we now focus on studying the $E_{p,i}-L_{p}$ relation of short bursts in statistics and try to explain its origin physically. Sample selection and data analysis are given in Section 2. The newly discovered $E_{p,i}-L_p$ relation is presented in Section 3. Implications of this relation are described in Section 4. We will end with conclusions and discussions in Section 5.

\section{Sample}
To study the $E_{p,i}-L_{p}$ relation of short bursts, the observed data of duration, redshift, peak flux and peak energy are basically necessary and should be available in advance. Seventeen short GRBs with measurements of $T_{90}$, $z$, $P$ and $E_{p,i}$ had been gathered and listed in Table 1, of which GRB 050709 and GRB 090510 are respectively detected by HETE-2 and Fermi, and others are detected by Swift/BAT. Among the 17 events, four special ``short'' bursts shown in Table 1 are taken from Ghirlanda et al. (2009).

Norris, Gehrels \& Scargle (2010) investigated the threshold effect on the detection of short GRBs with extended emission (EE) and found that the detection rate corrected for the physical threshold effect would be larger than the current rate of $\sim25\%$ estimated for the whole short burst sample. Meanwhile, short bursts with and without EE are thought to be originated from different progenitors (Norris, Gehrels \& Scargle 2011). The properties of short bursts with EE and their initial peaks are highly similar to those of the classical long and short bursts, respectively (Sakamoto et al. 2011). It happens that about one fourth of short bursts have the EE component in our short GRB sample in Table 1. The ratio is roughly the same as that of the whole sample of short bursts. People might feel puzzled on how to reclassify the short GRBs with EE in terms of their confusing observational features. Therefore, it is very interesting to compare classical short bursts with the short ones with EE via some empirical luminosity relations.

It needs to be noted that GRB 071020 and GRB 080913A in Table 1 are generally believed to be ``short'' burst despite of their prompt $\gamma$ duration longer that 2 seconds, since their spectral features in both the prompt and afterglow phase are very similar to those of short-hard ones (Greiner et al. 2009; Ghirlanda et al. 2009). The $T_{90}$ classification of GRBs is thus not very strict and some median bursts would not be clearly classified. In order to compare with previous $E_{p,i}-L_p$ relations of BATSE GRBs (Yonetoku et al. 2004) and Swift plus BATSE bursts (e.g. Wang, Qi \& Dai 2011), we have collected 148 long GRBs (110 Swift/BAT bursts, 10 Fermi/GBM bursts and 28 Konus-wind bursts) with the necessary quantities measured to construct a combined sample for getting the updated $E_{p,i}-L_{p}$ relation.

The observer-frame peak energy ($E_{p,o}$) and the rest-frame peak energy ($E_{p,i}$) are related by $E_{p,i}=E_{p,o}(1+z)$, where $E_{p,o}$ is derived as $E_{p,o}=(2+\alpha)E_0$ from $\nu F_{\nu}$ spectrum, of which $\alpha$ and $E_0$ are respectively lower energy index and break energy fitted with the Band function (Band 1993) for photon count spectrum in the observer frame. The observer-frame isotropic peak luminosity is determined by $L_p=4\pi d_l^2 P_{bolo}$, where the luminosity distance $d_l$ is $d_l=cH_0^{-1}(1+z)\int_{0}^z[\Omega_m(1+z')^3+\Omega_\Lambda]^{-1/2}dz'$ calculated on assumption of $H_0=71$ km s$^{-1}$ Mpc$^{-1}$, $\Omega_m=0.27$ and $\Omega_{\Lambda}$=0.73 (e.g. Dai, Liang \& Xu 2004). $P_{bolo}$ is the bolometric peak flux that is corrected for k-correction, $P_{bolo}=P\times K$, based on the observed peak flux $P$ (see Bloom, Frail \& Sari 2001; Amati et al. 2002; Schaefer 2007). We caution that the median $K$ value of 1.7 used by us for short bursts is slightly larger than the former value of $K\sim1$ for long bursts (Bloom, Frail \& Sari 2001). Furthermore, we have assigned a 10\% fluctuation as the uncertainties of $E_p$ and $P$ for several bursts whose measurement errors are unknown.


\section{The updated $E_{p,i}-L_p$ Relation}
Using 11 BATSE long bursts with known redshifts, Yonetoku et al. (2004) firstly fitted the $E_{p,i}-L_p$ relation with

\begin{equation}
\label{ep-lp}
\frac{L_{p}}{10^{52}ergs}=\kappa\times (\frac{E_{p,i}}{1keV})^{\nu},
\end{equation}
where the parameters $\kappa$ and $\nu$ are constrained as
\begin{equation}
\frac{L_p}{10^{52} ergs}=(2.34_{-1.76}^{+2.29})\times10^{-5}\times(\frac{E_{p,i}}{1 keV})^{2.0\pm0.2},
\end{equation}
from which one can draw a concise expression as $L_p\propto E_{p,i}^{2.0}$. Their result is consistent with an earlier plot shown by Wei \& Gao (2003). Ghirlanda et al. (2005) have re-studied this relation with a larger sample of 16 BATSE bursts and confirmed the relation of $L_p\propto E_{p,i}^{2.0}$. The index of $\nu\simeq2$ indicates the synchrotron radiation mechanism in a simple standard internal shock scenario is highly supported (Zhang \& M\'{e}sz\'{a}ros 2002; Wei \& Gao 2003; Rees \& M\'{e}sz\'{a}ros 2005). Very recently, Wang, Qi \& Dai (2011) analyzed a combined sample of 116 long GRBs including 31 pre-Swift bursts and 85 Swift/BAT bursts and derived the $E_{p,i}-L_p$ relation as
\begin{equation}
\frac{L_p}{10^{52} ergs}=(1.28_{-0.13}^{+0.17})\times(\frac{E_{p,i}}{300 keV})^{1.40\pm0.12},
\end{equation}
in which the expression of $L_p\propto E_{p,i}^{1.4}$ is obviously different from the so-called Yonetoku relation as mentioned above. In this work, we combine 148 long bursts with confirmed redshifts from Swift, Fermi and Konus data sets to retrieve the luminosity relation as
\begin{equation}
\frac{L_p}{10^{52} ergs}=(7.24_{-3.44}^{+6.56})\times10^{-5}\times(\frac{E_{p,i}}{1 keV})^{1.72\pm0.11},
\end{equation}
leading to $L_p\propto E_{p,i}^{1.7}$ with linear correlation coefficient of $R=0.79$ and a chance possibility less than 0.001, as shown in Figure 1. Our derived value of $\nu\simeq1.7$ is just located between 1.4 and 2.0, marginally consistent with Yonetoku's value within the 1$\sigma$ level. It is very important to identify which one should be more reliable because the underlying physics will emerge once the $\nu$ value is exactly decided as suggested by Rees \& M\'{e}sz\'{a}ros (2005).

On the other hand, the $E_{p,i}-L_p$ relation has not yet been obtained for short bursts due to the limited number of short GRBs with redshift measurements in the past decade. But now, 17 short bursts with measured redshifts are available to examine the possible luminosity relation. Hence, we can fit them with Eq. (1) for the first time in Figure 2 and present the new $E_{p,i}-L_p$ relation as follows,
\begin{equation}
\frac{L_p}{10^{52} ergs}=(1.07_{-0.91}^{+1.51})\times10^{-5}\times(\frac{E_{p,i}}{1 keV})^{1.73\pm0.44},
\end{equation}
where we get $L_p\propto E_{p,i}^{1.7}$ with linear correlation coefficient of $R=0.72$ and chance possibility of $P\sim0.001$, which is surprisingly in agreement with that of long bursts as shown in Eq. (4). In Figure 3, we compare the $E_{p,i}-L_p$ relation of short bursts with that of long ones and find that they are consistent with each other within 3$\sigma$ levels. It is very attractive that short GRBs with EE and ``long'' bursts with short-hard properties, compared with classical short bursts, have larger values of $E_{p,i}$ and $L_p$. Three kinds of short bursts populate in different regions on the $E_{p,i}-L_p$ plane. More interestingly, short bursts with EE reside at the high end of their $E_{p,i}-L_p$ relation, while ``long'' bursts with short-hard features seem to favor the $E_{p,i}-L_p$ relation of long bursts. In the following, we shall interpret the potential implication of the consistency of both kinds of bursts on the fundamental GRB physics.

As seen in Figure 4, the updated median values of redshift are 0.7 and 2.1 for short and long GRBs, respectively. The redshift distributions of both kinds of bursts show that they reside at different distances from us statistically. Although the spectral peak energies of short bursts are relatively harder than those of long bursts in the observer frame, both of the rest frame peak energies are very comparable (e.g. Zhang \& Choi 2008; Ghirlanda et al. 2012). In order to explore what causes the slight difference of parameter $\kappa$ in Eqs. (4) and (5), the luminosity distributions of 17 short and 148 long bursts are illustrated in Figure 5, from which we perform a gaussian fit to the $logL_p$ distributions and get the mean value of $\mu=52.24\pm0.03$ with a standard deviation of $\sigma=0.5$ ($\chi_\nu^2=2.4$) for long bursts and $\mu=52.07\pm0.18$ with a standard deviation of $\sigma=1.3$ ($\chi_\nu^2=0.5$) for short bursts. It is interesting to note that their median luminosities are nearly the same. However, a K-S test returns $D=0.29$ with a probability of 0.11, which is not significant enough for us to judge whether the two kinds of bursts are drawn from different parent populations. The similarity may be caused by the selection effect. For example, comparing with short bursts, long bursts usually have lower peak flux (Gehrels et al. 2008) but are more distant on average. So their mean luminosities are more or less similar.

\section{Implications}
Rees \& M\'{e}sz\'{a}ros (2005) suggested that a laminar and relativistic jet as seen head on will produce a (probably Comptonized) thermal spectrum peaking at the hard X-ray or $\gamma$-ray energy bands. They also pointed out that the dependence of spectral peak energy of GRB parameters relies on the specific radiation mechanisms forming the spectrum in those energy bands. According to diverse assumptions of the emission mechanism, they proposed three spectral cases that gave rise to the corresponding $E_{p,i}-L_p$ relation as
\begin{eqnarray}
E_{p,i}\propto    
\left\{                      
\begin{array}{lll}       
\Gamma^{-2}t_{var}^{-1}L^{1/2}& (Case 1: synchrotron \ radiation)\\  
  r_0^{-1/2}L^{1/4}\sim L^{(2\beta'+1)/4}& (Case 2: thermal\ radiation)\\
 \Gamma^2L^{-1/4}\sim L^{(8\beta-1)/4} & (Case 3: Comptonization)\\
\end{array}              
\right.                       
\end{eqnarray}
where $L =L_p$, $t_{var}$ is the typical variability timescale of the outflow, $\Gamma$ and $r_{0}$ are Lorentz factor of ejecta and typical radius of photosphere from central engine, respectively. Note that the scaling laws of Cases 1 and 2 have been derived as early as in 2002 by Zhang \& M\'{e}sz\'{a}ros (2002). The parameters $\beta$ and $\beta^{'}$ are two power-law indices in their assumed relations of $\Gamma\propto L^{\beta}$ and $r_{0}\propto L^{-\beta'}$. For $\beta'=(0.5, 1)$, one can easily obtain  $E_{p,i}\propto(L^{1/2}, L^{3/4})$. Because the observable half-opening angle satisfies $\theta\sim1/\Gamma$ and $L\propto\theta^{-2}$ (Frail et al. 2001), we can get $\Gamma\propto L^{1/2} (\beta=1/2)$ and then $E_{p,i}\propto L^{3/4}$ for the case 3 in Eq. (6).

Comparing the results listed in Eqs. (2)-(5) with Eq. (6), we find that Yonetoku's result is consistent with case 1 while Wang et al's result supports case 3. However, our results in Eqs.(4) and (5) indicating the form of $E_{p,i}\propto L^{0.6}$ are coincident with case 2, from which one can notice that the index 0.6 is just in the range from 1/2 to 3/4. Meanwhile, the parameter $\beta'$ in Eq. (6) is then determined to be $\beta'=0.68$, also within (0.5, 1). The consistent power-law index highly hints that the spectrum of long and short bursts could be resulted from the same mechanism (see also Ghirlanda et al. 2011), whose dominant emission is thermal (also Fan et al. 2012), rather than synchrotron or Comptonization. It implies that the single synchrotron mechanism combined with the standard internal shock model is not sufficient to account for all GRB phenomena, especially the $E_{p,i}-L_p$ relation.

\section{Discussions and Conclusions}
We investigated the relations of the peak luminosity with the peak energy for long and short gamma-ray bursts in the new-satellite era. We found that a consistent correlation of $L_{p}\propto E_{p,i} ^{1.7}$ coexisted in both long and short GRBs, which provided a promising evidence that two kinds of bursts should be the produced from the same radiation process. Some key parameters of the process are constrained and suggest that the radiation process of both long and short bursts may be dominated by thermal emission, rather than the single synchrotron radiation. Our results challenge the theoretical models.

It needs to be emphasized that the index $\nu$ of $E_{p,i}-L_p$ relation in the Swift/Fermi era is systematically smaller than that in the BATSE era. Most probably, this phenomenon might be due to the different flux threshold as described in Wei \& Gao (2003). Moreover, the fact that Swift/BAT is more sensitive to softer and longer GRBs than BATSE (Band 2006; Gehrels et al. 2008) can inevitably lead to a distinguishing measurement of peak luminosity. In virtue of the lower flux threshold, Swift/BAT can detect very faint bursts which may contribute to the updated $E_{p,i}-L_p$ relation differently. Meanwhile, Swift and HETE-2 are lower and narrower than BATSE in energy ranges of detectors, which causes the spectral peak energies observed with Swift and HETE-2 to be systematically lower. In addition to the higher sensitivity of Swift/BAT, spectral redshifting and cosmological time dilation move high-redshift bursts into the parameter region where BAT is more sensitive than BATSE. In a sense, the systematically larger redshifts measured by Swift might also affect the power law index of the $E_{p,i}-L_p$ relation.

In general, short and long bursts are believed to have different distributions of observed or intrinsic durations, redshifts, spectral hardness, energy injections, circum-burst environments, and prompt $\gamma$-ray emitting regions (Zhang et al. 2008, 2011), etc, indicating their diverse origins (e.g. Piran 2004; Zhang 2007; Gehrels et al. 2009 for reviews). However, recent contrastive investigations demonstrate that two classes of bursts are most probably produced by the same radiation mechanism regardless of their different progenitors (e.g. Lee \& Ramirez-Ruiz 2007). For instance, Hakkila \& Preece (2011) studied the distinguishable pulses of short and long GRBs and found some common properties for correlations between duration, peak luminosity, spectral hardness, energy-dependent lag, and asymmetry (see also Zhang 2007). Furthermore, Boci et al. (2010) argued that the correlative pulse relations cannot be obtained as the direct consequence of the synchrotron emission in the framework of the standard relativistic shock model. All these observational evidences seem to validate that short and long bursts might be originated from the same radiation mechanism independent of their different progenitors.

Based on the analysis and discussions mentioned above, we draw our conclusions as follows: (1) Long bursts in the Swift era follows the $L_P\propto E_{p,i}^{1.7}$ relation different from that of $L_P\propto E_{p,i}^{2.0}$ for the BATSE GRB sample. The difference may be caused by the threshold effect of different detectors; (2) The relation between $L_P$ and $ E_{p,i}$ for 17 short bursts is found to be $L_P\propto E_{p,i}^{1.7}$ that agrees well with that of long bursts in the Swift era, except the minor difference in intercept of logarithmic luminosity; (3) The spectrum of short and long GRBs could be produced by the same radiation mechanism possibly dominated by the thermal component, other than the solely synchrotron or Comptonized process. This would greatly challenge the current theories of both kind of GRBs.


\acknowledgments
We thank the anonymous referee for her/his critical and helpful comments. This work is supported by the National Natural Science Foundation of China (Grant No. 10943006), Guizhou natural and scientific foundations (Grant No. 20092662, 20104014, 20117006 and 20090130) and foundations of Ministry of Education of the PRC (No. 210197, 20101174). Y.F.H. was supported by the National Natural Science Foundation of China (Grant No. 11033002) and the National Basic Research Program of
China (973 Program, Grant No. 2009CB824800).


%
%
%



\clearpage




%

\clearpage




%
 \begin{figure}
 \includegraphics[angle=0,scale=1.5]{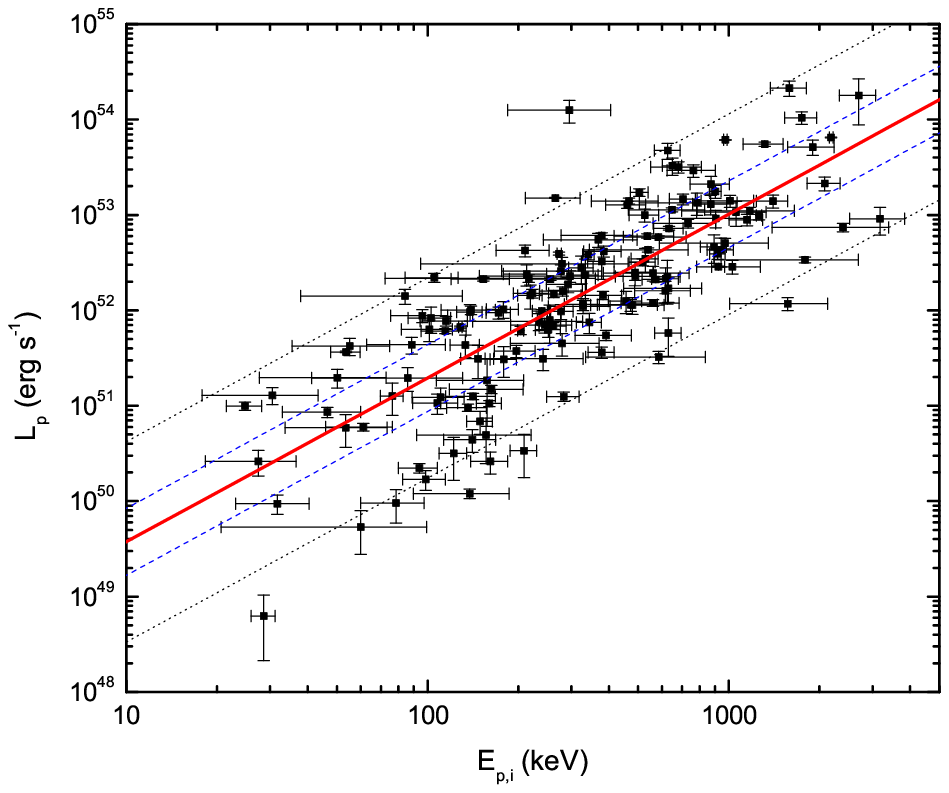}
      \caption{Rest frame peak energy versus peak luminosity for 148 long GRBs with measured redshifts. The solid line is the best fit to the observed data in the logarithmic frame of axes. The dashed lines and the dotted lines correspond to its 1$\sigma$ and 3$\sigma$ scatters ($\sigma =0.35$), respectively. }
         \label{Ep-LpofLB}
   \end{figure}
\clearpage
 \begin{figure}
 \includegraphics[angle=0,scale=1.5]{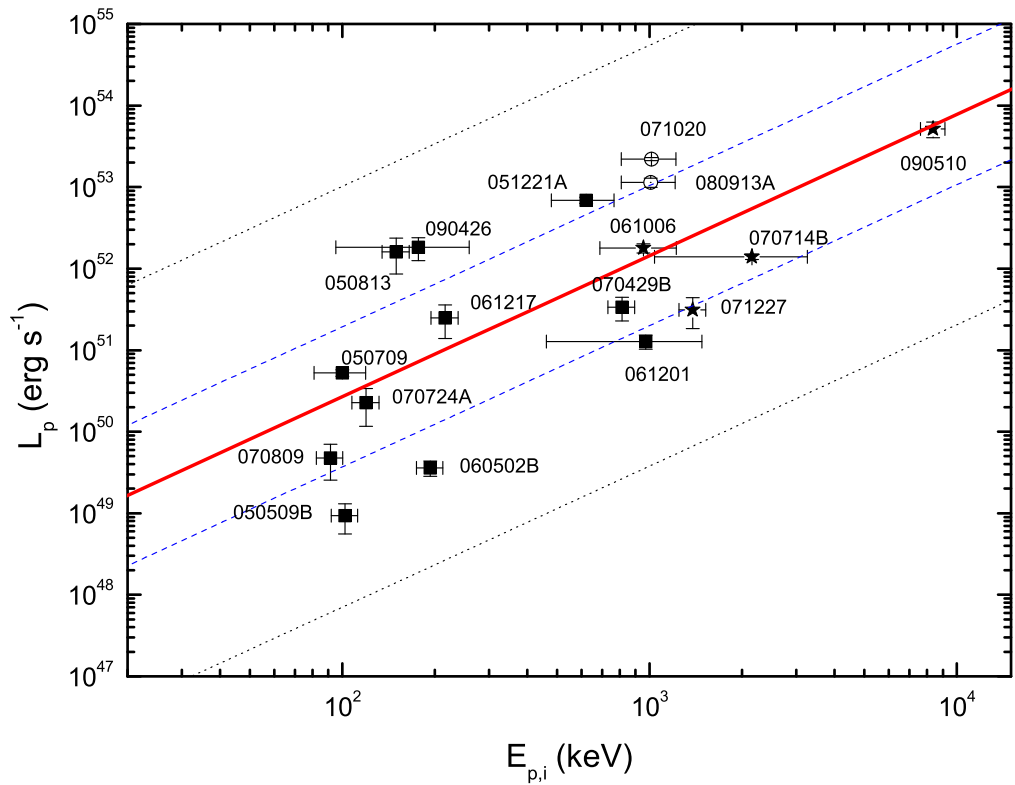}
      \caption{Rest frame peak energy versus peak luminosity for 17 short GRBs with measured redshifts. The solid line is the best fit to the observed data in the logarithmic frame of axes. The dashed lines and the dotted lines correspond to its 1$\sigma$ and 3$\sigma$ scatters ($\sigma=0.86$), respectively. ``Short'' GRBs with EE are symbolized with dark filled stars and ``long'' bursts with short-hard properties are marked with open circles. }
         \label{Ep-LpofSB}
   \end{figure}
\clearpage
 \begin{figure}
 \includegraphics[angle=0,scale=1.5]{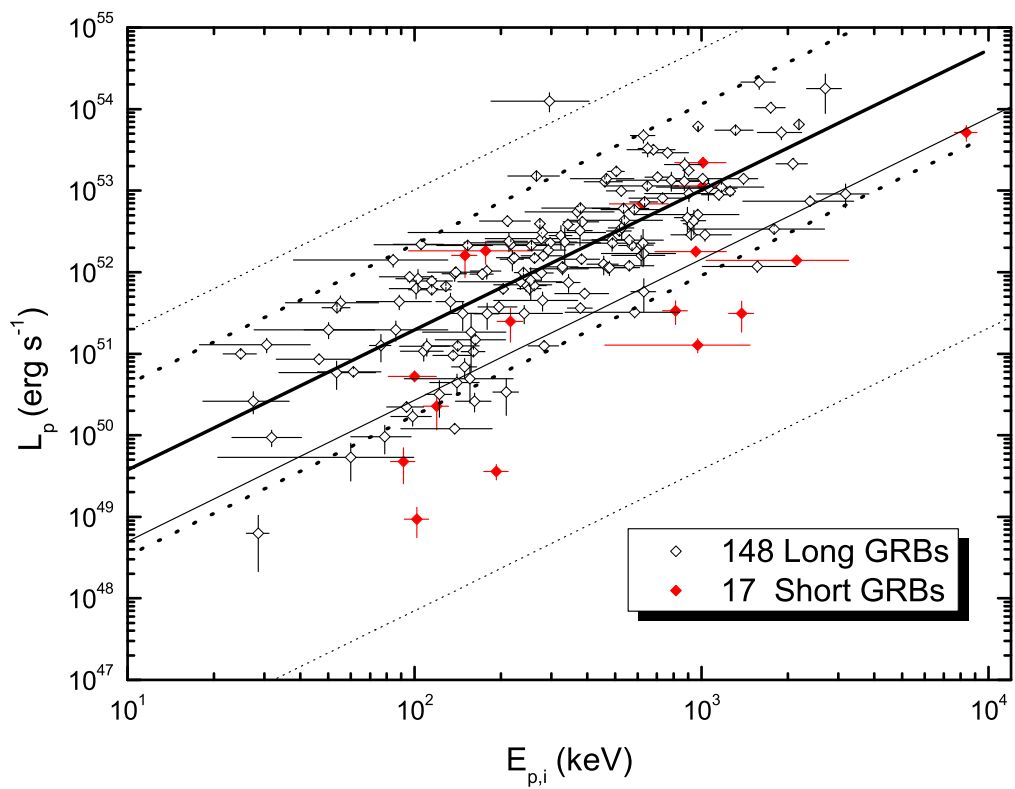}
      \caption{Comparison of rest frame peak energy versus peak luminosity for 17 short (filled diamond) and 148 Swift long (empty diamond) GRBs. The best fit lines (solid) and their 3$\sigma$ levels (dotted) are respectively taken from Figs. 1 and 2. They are symbolized by the thick line and the thin line, individually.}
         \label{Ep-LpofLandSB}
   \end{figure}

   \begin{figure}
   \centering
 \includegraphics[angle=0,scale=1.5]{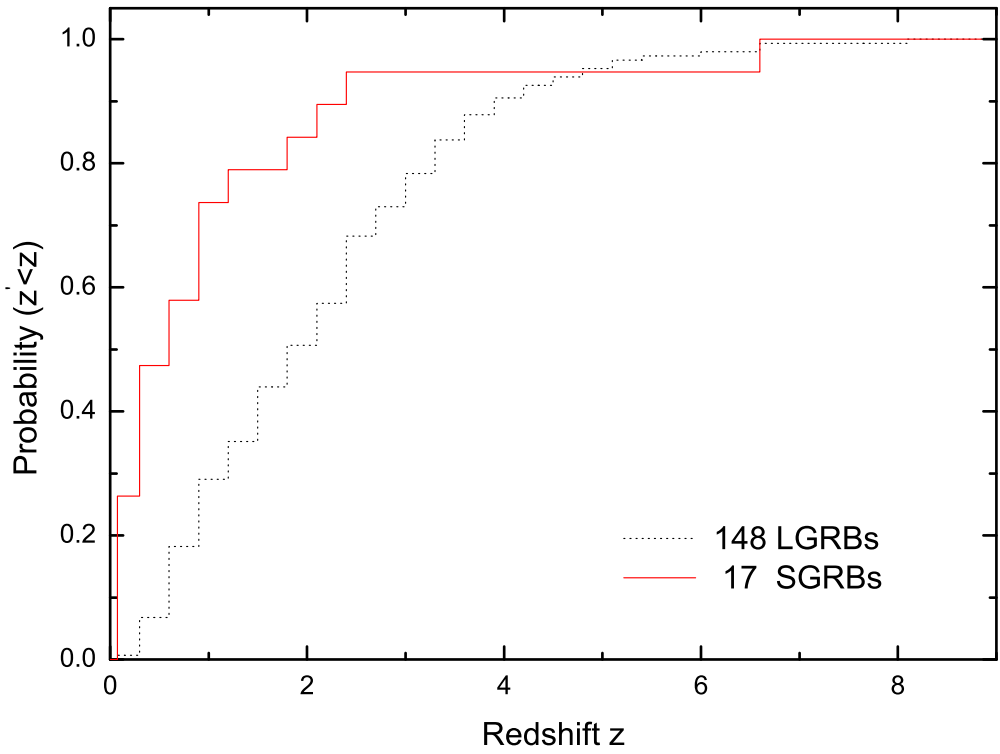}
      \caption{ Cumulative redshift distributions of 148 long and 17 short GRBs are displayed with dotted and solid lines, respectively.  }
         \label{beta-gamma}
   \end{figure}

\begin{figure}
   \centering
 \includegraphics[angle=0,scale=1.5]{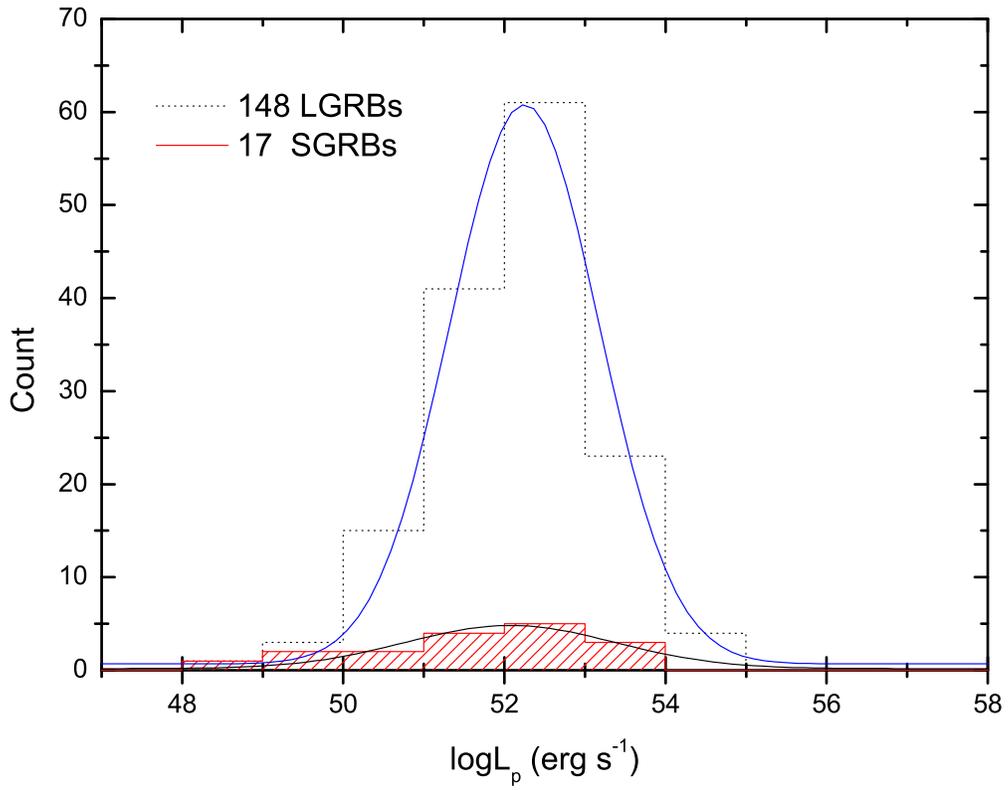}
      \caption{The histogram distributions of logarithmic luminosity are presented for 148 long (dotted line) and 17 short (solid line) GRBs. They are fitted by Gaussian functions, respectively.}
         \label{gamma-theta}
   \end{figure}





%
\clearpage

\begin{deluxetable}{ccrcrrrrr}
\tabletypesize{\scriptsize}
\tablecaption{A Sample (N=17) of Short GRBs with Measured Redshifts }
\tablewidth{0pt}
\tablehead{
\colhead{GRB$^\star$} & \colhead{$T_{90}$ } & \colhead{z} & \colhead{$P^{\sharp}$ }&Range & \colhead{$L_p$ } &
 $K^{\dag}$& \colhead{$E_{p,i}^{\sharp}$ }  & \colhead{Refs}\\
&(seconds)&&($10^{-6}$ erg/cm$^{2}$/s)&keV&($10^{51}$ erg/s)&&(keV) &
}
\startdata
050509B& 0.04& 0.225&0.51$\pm$0.3 &15-350  & 0.009$\pm$0.004& 2.8 & 102$\pm$[10]& (6,6,5,5)\\
050709&0.07 & 0.16&5.1$\pm$0.5 &2-400 & 0.53$\pm$0.05& 1.5&  100$\pm$19& (7,2,2,7) \\
050813&0.6&1.8 & 0.41$\pm$0.19 &15-350 & 16.07$\pm$7.5& 1.7&  150$\pm$[15]  & (6,6,5,5)\\
051221A&1.4 &0.547 &46$\pm$13 &20-2000&69.1$\pm$ 6& 1.2& 621$\pm$144 &(6,6,5,5) \\
060502B& 0.131& 0.287&1.89$\pm$1.49&15-350 &0.04$\pm$0.01 &2.7 &193$\pm$[19] & (6,6,5,5)\\
061006$^b$&1 &0.4377 & 21$\pm$2 &20-2000&  17.8$\pm$2.3&1& 955$\pm$267&(1,2,2,2) \\
061201& 0.76& 0.111&2.45$\pm$1.95 &15-350&1.27$\pm$0.25 &1.3 & 969$\pm$508& (6,6,5,5)\\
061217& 0.21& 0.827&0.44$\pm$0.2&15-350& 2.49$\pm$1.1&1.7 & 216$\pm$[22] &(6,6,5,5) \\
070429B&0.47 &0.904 & 0.43$\pm$0.14&15-350& 3.37$\pm$1.08& 1.9& 813$\pm$[81] & (6,6,5,5)\\
070714B$^b$&2 & 0.92&1.2$\pm$0.9 &15-350& 14$\pm$1& 1& 2150$\pm$1113&(3,2,2,2) \\
070724A& 0.4& 0.457& 0.14$\pm$0.06&15-350&0.23$\pm$0.11 &2.2 & 119$\pm$12 & (6,6,5,5)\\
070809&1.3 &0.2187 &0.17$\pm$0.08&15-350 &0.05$\pm$0.02& 1.9&91$\pm$[9]  & (6,6,5,5)\\
071020$^{a}$&4 &2.145 & 6$\pm$0.6 &20-2000&220$\pm$10 &1 &1013$\pm$205 &(4,2,2,2) \\
071227$^{b}$& 1.8& 0.383&3.5$\pm$1.1&20-1300 &3.11$\pm1.28$ & 1.8& 1383$\pm$[138] & (6,6,5,5)\\
080913A$^{a}$& 8& 6.7& 0.13$\pm$0.08&15-350&  114$\pm$15&1 &1009$\pm$200 &(4,2,2,2) \\
090426& 1.2& 2.609&0.34$\pm$0.23&15-350 & 18.22$\pm$5.7& 2.1& 177$\pm$82 &(6,6,5,5) \\
090510$^{b}$& 0.3&0.903 & 40$\pm$[4]&8-40000& 516$\pm$112& 0.6& 8373$\pm$761 & (6,6,8,8)\\
\enddata
\tablecomments{References are given in order for duration, redshift, peak flux and rest frame peak energy, respectively.
1. Hurley et al. 2006, GCN 5702; 2. Ghirlanda et al. 2009; 3 Kodaka et al. 2007, GCN 6637; 4. Greiner et al. 2009; 5. Butler et al. 2007 (http://butler.lab.asu.edu/swift/); 6. http://heasarc.nasa.gov/docs/swift/swiftsc.html; 7. Villasenor et al. 2005; 8. Ackermann et al. 2010.}
\tablenotetext{a} {\ ``Long'' GRBs with short-hard properties (see the text in section 2).}
\tablenotetext{b} {\ ``Short'' GRBs with extended emission. GRB 090510 is taken from Abdo et al. (2009), others are drawn from Norris, Gehrels \& Scargle (2010).}
\tablenotetext{\star}{\ GRBs 051221A, 061201 and 071227 are discovered by Konus-wind; GRBs 090510 and 050709 are detected by Fermi/GBM and HETE-2, respectively; The other 12 bursts are observed by Swift/BAT.}
\tablenotetext{\sharp}{\ Values without measured errors have been given a 10\% fluctuation, as shown by square brackets.}
\tablenotetext{\dag}{\ The value of K=1 has been assigned to the four GRBs selected from Ghirlanda et al. (2009).}
\end{deluxetable}
%
\clearpage 

\end{document}